\documentclass[sigconf]{acmart}
\renewcommand\footnotetextcopyrightpermission[1]{} % removes footnote with conference information in first column

\usepackage[font=small,labelfont=normalfont,textfont=md]{caption}
\usepackage{fancyvrb}
\usepackage{setspace}
\usepackage{url}
\usepackage{color}
\usepackage{paralist}
\usepackage[small,compact]{titlesec}
\usepackage{setspace}
\usepackage{times}
\usepackage{paralist}
\usepackage{amsmath,amssymb,amsfonts,amsthm}
\usepackage[geometry ]{ifsym}
\usepackage[T1]{fontenc}
\usepackage[latin9]{inputenc}
\usepackage{listings}
\usepackage{nicefrac}
\lstdefinestyle{nonumbering}{
        basicstyle=\ttfamily\scriptsize,
        language=Java,
        emph={},
        emphstyle={\underbar},
        numbers=none,
        numberfirstline=false,
        numberstyle=\tiny,
        escapeinside={(*@}{@*)}
}
\lstdefinestyle{small}{
        basicstyle=\tiny,
        language=Java,
        emph={},
        emphstyle={\underbar},
        numbers=none,
        numberfirstline=false,
        numberstyle=\tiny,
        escapeinside={(*@}{@*)}
}

\usepackage{hyperref}

\usepackage{float}
\usepackage{graphics}
\usepackage{graphicx}
\usepackage{wrapfig}
\usepackage{epsfig}
\usepackage[linesnumbered,ruled,vlined]{algorithm2e}
\usepackage{rotating}
\usepackage{mathpartir}
\usepackage{multirow}
\usepackage{booktabs}
\usepackage{tabularx}
\usepackage{cleveref}

\usepackage{ifthen}

\newcommand{\DONE}[1]{}
\newcommand{\COMMENT}[1]{}

\newcommand{\figref}[1]{Fig.~\ref{Fi:#1}}

\newcommand{\tabref}[1]{Tab.~\ref{Ta:#1}}
\newcommand{\secref}[1]{Section~\ref{Se:#1}}

\newcommand{\figlabel}[1]{\label{Fi:#1}}

\newcommand{\tablabel}[1]{\label{Ta:#1}}
\newcommand{\seclabel}[1]{\label{Se:#1}}

\newtheorem{Example}{Example}
\newcommand{\ignore}[1]{}

\newboolean{TR}
\setboolean{TR}{false}
\ifthenelse{\boolean{TR}}{

\newcommand{\TrOnly}[1]{#1}
\newcommand{\SubOnly}[1]{}
\newcommand{\TrOnlyInFootnote}[1]{#1}
\newcommand{\TrOnlyInTable}[1]{#1}}
{

\newcommand{\TrOnly}[1]{}
\newcommand{\SubOnly}[1]{#1}
\newcommand{\TrOnlyInFootnote}[1]{}
\newcommand{\TrOnlyInTable}[1]{}}

\newtheorem{claim}[theorem]{Claim}

\newcommand{\sectionette}[1]{\noindent \textit{#1}:}

\newcommand{\pvaleq}[1]{p\!\!=\!\!#1}

\newcommand{\hiddentext}[1]{}

\newcommand{\para}[1]{\vspace{3pt}\noindent\textbf{\textit{#1}}}
\newcommand{\scode}[1]{{\small \texttt{#1}}}

\renewcommand{\phi}{\varphi}

\newcommand{\vocab}{\mathcal{V}}

\newcommand{\examples}{\mathcal{E}}

\newcommand{\affix}{\textit{affix}}

\newcommand{\retain}{\textit{retain}}

\newcommand{\exclude}{\textit{remove}}

\newcommand{\sem}[1]{\llbracket #1 \rrbracket}

\acmConference{September}
\acmYear{2017}
\acmMonth{10}
\startPage{1}
\begin{document}

\setlength{\pdfpageheight}{\paperheight}
\setlength{\pdfpagewidth}{\paperwidth}

\title{Programming Not Only by Example}
\author{Hila Peleg}
\affiliation{Technion}
\email{hilap@cs.technion.ac.il}
\author{Sharon Shoham}
\affiliation{Tel Aviv University}
\email{sharon.shoham@gmail.com}
\author{Eran Yahav}
\affiliation{Technion}
\email{yahave@cs.technion.ac.il}

\begin{abstract}

In recent years, there has been tremendous progress in automated synthesis techniques that are able to automatically generate code based on some intent expressed by the programmer. A major challenge for the adoption of synthesis remains in having the programmer communicate their intent. When the expressed intent is coarse-grained (for example, restriction on the expected type of an expression), the synthesizer often produces a long list of results for the programmer to choose from, shifting the heavy-lifting to the user. An alternative approach, successfully used in end-user synthesis is programming by example (PBE), where the user leverages examples to interactively and iteratively refine the intent. However, using only examples is not expressive enough for programmers, who can observe the generated program and refine the intent by directly relating to parts of the generated program.

We present a novel approach to interacting with a synthesizer using a granular interaction model. Our approach employs a rich interaction model where
\begin{inparaenum}[(i)]
\item the synthesizer decorates a candidate program with debug information that assists in understanding the program and identifying good or bad parts, and
\item  the user is allowed to provide feedback not only on the expected output of a program, but also on the underlying program itself. That is, when the user identifies a program as (partially) correct or incorrect, %they can also provide a justification for their decision. The justification allows the user to accept or discard parts of the program.
    they can also explicitly indicate the good or bad parts, to allow the synthesizer to accept or discard parts of the program instead of discarding the program as a whole.
\end{inparaenum}

We show the value of our approach in a controlled user study. Our study shows that participants have strong preference to using granular feedback instead of examples, and are able to provide granular feedback much faster.

\end{abstract} 
\maketitle
\section{Introduction}
%Synthesis works (!)

In a development ecosystem where programmers are frequently asked to take on tasks involving unfamiliar APIs and complex data transformations, program synthesis is both a tool to shorten development times and an aid to small tasks of API programming.

Synthesis tools for end-users are available for a wide variety of purposes from creating formulae in Microsoft Excel~\cite{Gulwani:2011:ASP:1926385.1926423} to formulating SQL queries~\cite{wang2017synthesizing}. Tools for expert users who can encode full specifications have also matured enough to be practical~\cite{solar2008program,liu2012dynamic,VYY:POPL10}.

\para{Expressing Intent}
Despite significant progress in synthesis, expressing the user's intent remains a major challenge. An expert user can write full specifications and express their intent fully~\cite{liu2012dynamic,VYY:POPL10,Blaine:1998:PDS:521138.786844,Itzhaky:2010:SIS:1932682.1869463,paige1990symbolic,Chowdhury:2016:AAD:2851141.2851167,shachar2016,vechev2006correctness,vechev2007cgcexplorer,hou2011evaluation}, but end-users and programmers trying to solve small tasks often use \emph{partial specifications}. Partial specifications are available in different forms, depending on the synthesizer: source and target types, input-output pairs, tests, and logical specifications.

Coarse-grained models such as type-driven synthesis present the user with all possible results that satisfy the coarse-grained criteria (e.g., \cite{gvero2013complete,galenson2014codehint}). This leads to a challenging task: %comparing a large number of similar programs to select one that solves the user's problem.
the user must compare a large number of similar programs to select a solution.

%A common practice in end-user synthesis is to leverage user-provided examples to interactively and iteratively refine the intent (Petri-nets, tests, etc)
\para{Expressing Intent with Examples}
An alternative that has proven itself extremely useful for end-users is to use examples as a way to express intent. Programming by Example (PBE) is a form of program synthesis where the desired behavior is generalized from specific instances of behavior, most often input-output example pairs. This allows an iterative process where, if the synthesized program is not acceptable, additional examples are provided until the target program is reached. This technique is often used either on its own in synthesizers such as \cite{Gulwani:2011:ASP:1926385.1926423,wang2017synthesizing,lau2001learning,landauer1995visual,witten1993tels,anton2005xpath,wu2015automatic,OmariSY16icse,PLDI-2014-LeG} or as a way to refine the results of type-driven synthesis~\cite{osera2015type,feser2015synthesizing}.

\para{Insufficiency of Examples for Programmers} PBE is geared towards end-users, but is also useful for more advanced users when the behavior is more difficult to describe than its effect. However, in this interaction model, a user can only do one of two things: accept the program after inspection, or reject it with a \emph{differentiating example} which will rule it out in the next iteration of synthesis. There is wasted knowledge in forcing a programmer to work within this interaction model, as some synthesized programs are not \emph{all bad} - they may have a part of the program that is overfitted to the examples, while another will be on the right track. Allowing only a full accept or full reject ignores the ability of a programmer to read and understand the program, and to express a more directed, \emph{granular} feedback, deeming parts of it as desirable or undesirable, rather than the program as a whole.

%\para{Hypothesis} We hypothesize that,
In fact, we hypothesize that, in some cases, it is \emph{easier} for a programmer to explicitly indicate what is good or bad in a candidate program, instead of implicitly trying to express this information through input-output examples.
Moreover, we prove that it is sometimes \emph{impossible} to express such information through examples.

\para{Programming Not Only by Example}
Motivated by the insufficiency of examples, we present a new, \emph{granular interaction model} that allows a programmer to interact with the synthesizer \emph{not only by example} but also provide feedback on parts of the synthesized program. Our interaction model is granular in both directions, from the programmer to the synthesizer, and back:

\sectionette{Synthesizer $\rightarrow$ Programmer}
A candidate program is presented together with debug information, showing execution values at different program points. This helps the programmer understand whether the candidate program behaves as expected at \emph{intermediate states}, instead of relying only on its final output.

\sectionette{Programmer $\rightarrow$ Synthesizer}  A programmer can provide:
\begin{inparaenum}[(i)]
\item input-output examples (as in PBE), and
\item \emph{granular} feedback on the candidate program by explicitly accepting/rejecting \emph{parts} of its code.
\end{inparaenum}

%The motivation for this form of interaction is the observation that synthesis with input-output examples requires the user to provide examples that differentiate the candidate program from the desired program. Our hypothesis is that, in some cases, it is easier for a programmer to explicitly indicate what is good or bad in a candidate program, instead of implicitly trying to express this information through input-output examples. This hypothesis is strongly supported by a controlled user-study with $32$ developers from both academia and industry. To conduct this study, we developed a synthesizer that interacts with the user in three different ways: holistic (PBE), granular, or both. Our synthesizer also measures interaction times, and records the user-interaction so we can later analyze it.

We tested the granular interaction model by a controlled user-study with $32$ developers from both academia and industry. To conduct this study, we developed a synthesizer that interacts with the user in three different ways: holistic (PBE), granular, or both. Our synthesizer also measures interaction times, and records the user-interaction so we can later analyze it.

Our implementation synthesizes functional programs in Scala. Scala is a popular functional and object-oriented programming language, used in many big-data processing frameworks (e.g., Spark, Akka). Functional compositions are considered ``the Scala way'' to approach coding tasks, and so we aim to synthesize them.

\para{Advantages of granular interaction}
The user study strongly supports the hypothesis that it is beneficial to let programmers communicate their understanding of the program \emph{explicitly} to the synthesizer (by marking parts of it as desirable or undesirable) rather than \emph{implicitly} (through examples). Several participants in our user study, faced with the inability to rule out an undesired operation in the program using only examples, expressed extreme frustration.
%Furthermore, the study also demonstrates the inability to rule out some undesired operations with examples only.
As we show in our user-study, this is more common than one would imagine, due to the introduction of redundant or superfluous operations by the synthesizer. As a result, an undesirable operation may be part of several candidate programs along the process, but the holistic PBE model does not allow ruling it out.

%in some cases, when the requirements from the desired program are non-functional, \emph{it is impossible to eliminate certain bad programs} which are functionally equivalent to the desired program. As we show in our user-study, this is more common than one would imagine, due to the introduction of redundant or superfluous operations by the synthesizer. A similar problem sometimes occurs before an equivalent program is reached: an undesirable operation may be part of several candidate programs along the process, but the holistic PBE model does not allow ruling it out.

Our study shows that our granular interaction model (GIM) is easier to use, as supported by:
\begin{inparaenum}[(i)]
\item a strong preference of participants for granular feedback over examples, and
\item a significantly shorter iteration time when using granular feedback.
\end{inparaenum}
It is important to note that granular feedback does not completely replace examples. Participants that were restricted to granular feedback were sometimes forced to use a larger number of iterations, and were more prone to error when accepting the program. We therefore conclude that future synthesizers should integrate both interaction models.

\para{Main Contributions}
The contributions of this paper are:
\begin{itemize}
 \item A synthesis framework for programmers with a  granular interaction model (GIM), which allows the user to approve or disapprove of specific parts of the code of the candidate program, rather than just respond to it as a whole; and allows a synthesizer to present candidate programs with debug information.
 \item A theoretical result that shows that examples are sometimes insufficient for reaching the desired program. We further show that this insufficiency occurs in practice throughout real PBE sessions.
 %\TODO{A formalization of user-driven interactive synthesis}, which allows us to crystalize the shortcomings of working with the holistic PBE model,and to show how our interaction model complements it.
  \item A controlled user study showing that programmers have strong preference for granular feedback instead of examples, and are able to provide granular feedback much faster.
\end{itemize}

\paragraph{Outline}
\Cref{insufficient} shows why examples are not only inconvenient but insufficient to communicate the intent of the programmer. To allow more expressive power, we introduce three additional granular operations in  \secref{granular}. Our full interaction model consists of the granular operations
% which were demonstrated above
 as well as examples.
In addition, we also introduce debug information for every example provided by the user in \secref{debugview}.

In \secref{experiments} we detail our experiments on the number of iterations necessary to solve a set of benchmarks with different interaction models. We also detail a controlled user study of $32$ programmers from academia and industry.
The study shows an advantage of granular predicates over examples in both iteration time and preference, and we discuss the need for both the granular predicates and examples in order to help the user reach the (correct) target program.

\section{Overview}\seclabel{overview}

In this section we provide an overview of our Granular Interaction Model (GIM) for synthesis on a simple example.
We start by showing the interaction model of Programming by Example (PBE) and its shortcomings, and then describe how GIM overcomes these shortcomings by using a richer interaction model.

\paragraph{Motivating example}\seclabel{mot-example}

Consider the task of writing a program that \emph{finds the most frequent character-bigram in a string}. Assume that the program is constructed by combining operations from a predefined set of operations we refer to as the vocabulary $\vocab$. For now, assume that the vocabulary contains standard operations on strings, characters, and lists.
In addition, assume that you provide an initial \emph{partial specification} in the form of an input-output example: \[\sigma_0=\scode{"a\underline{bd}fibfcfde\underline{bd}fde\underline{bd}ihgfkjfde\underline{bd}"}\mapsto\scode{"bd"}.\]

In this example, the bigram ``bd'' is the most frequent (appears $4$ times), and is thus the expected output of the synthesized program.

\ignore{
\paragraph{The space of possible solutions}
When the vocabulary for synthesis goes beyond a carefully designed domain-specific language, even a small vocabulary can lead to a huge number of programs that satisfy the partial specification at each step of synthesis~\cite{polozov2015flashmeta}. In our example, there are $674$ programs satisfying the initial specification, out of a search space of  $118,261$ programs (considering only programs up to length $6$). User interaction is extremely important as a way to guide the search in this large space.
}

\begin{table}[t]
  \centering \small
    \begin{tabular}{ll}
  \multicolumn{2}{l}{\textbf{Task:} find the most frequent bigram in a string} \\
  \hline
%  Step  & PBE \\
%  \hline
  Initial example ($\sigma_0$) & \scode{"abdfibfcfdebdfdebdihgfkjfdebd"}$\mapsto$\scode{"bd"}\\
  \hline
  Question $q_1$ & \begin{lstlisting}
input
.takeRight(2)\end{lstlisting} \\
\hline
\multicolumn{2}{l}{\textbf{Problem:} \scode{takeRight} will just take the right of a given string}\\
\multicolumn{2}{l}{\textbf{Idea:} the frequent bigram needs to be placed in the middle}
\\
  \hline
  Answer $\sigma_1$ & \scode{"cababc"}$\mapsto$\scode{"ab"} \\
  \hline
  \begin{tabular}{c}Question \\ $q_2$\end{tabular}& \begin{lstlisting}
input
.drop(1)
.take(2)
  \end{lstlisting}\\
\hline
\multicolumn{2}{l}{\textbf{Problem:} this program crops a given input at a constant position}\\
\multicolumn{2}{l}{\textbf{Idea:} vary the position of the frequent bigram between examples}\\
\hline
Answer $\sigma_2$  & \scode{"bcaaab"}$\mapsto$\scode{"aa"} \\
\hline
\begin{tabular}{c}Question \\ $q_3$\end{tabular}& \begin{lstlisting}
input
.zip(input.tail)
.drop(1)
.map(p => p._1.toString + p._2)
.min
\end{lstlisting}\\
\hline
\multicolumn{2}{l}{\textbf{Problem:} in all examples the output is the lexicographical minimum of all}\\
\multicolumn{2}{l}{bigrams in the string (e.g., "aa" < "bc", "aa" < ca", "aa" < "ab")}\\
\multicolumn{2}{l}{\textbf{Idea:} have a frequent bigram that is large in lexicographic order}\\

\hline
Answer $\sigma_3$  & \scode{"xyzzzy"}$\mapsto$\scode{"zz"} \\
\hline
%$\vdots$\\
    \end{tabular}%
  %\caption{Handling programs overfitted to the examples with more examples}\label{overview-walkthrough}%
  \caption{The difficulty of finding a differentiating example.}\label{pbe-walkthrough}
\end{table}%

\subsection{Interaction with a classical PBE synthesizer}\seclabel{overview-pbe}

\Cref{pbe-walkthrough} shows the interaction of a programmer with a PBE synthesizer to complete our task. The synthesizer poses a question to the programmer: \emph{a candidate program that is consistent with all examples}. The programmer provides an answer in the form of an \emph{accept}, or additional \emph{input-output examples} to refine the result.%\TODO{Switched the order, I think it's more readable}%, or a decision to accept the synthesized program.

Based on the initial example, the synthesizer offers the candidate program $q_1$, which consists of a single method from the vocabulary --  \scode{takeRight(2)}, which returns the $2$ rightmost characters -- applied to the input. The programmer then responds by providing the example $\sigma_1$ which is inconsistent with the candidate program, and therefore \emph{differentiates} it from the target program.

At this point, the synthesizer offers a new candidate program $q_2$ that is consistent with both $\sigma_0$ and $\sigma_1$.

The interaction proceeds in a similar manner.
Each additional example may reduce the number of candidate programs (as they are required to satisfy all examples). With a careful choice of examples by the user, the process terminates after a total of $4$ examples.

\paragraph{Finding differentiating examples may be hard} Consider the candidate program $q_3$. To make progress, the user has to provide an example that differentiates $q_3$ from the behavior of the desired program. To find a differentiating example, the user must
\begin{inparaenum}[(i)]
\item understand the program $q_3$ and why it is wrong, and
\item provide input-output examples that overrule $q_3$, and preferably also similar programs.
\end{inparaenum}

By examining the code of $q_3$, it is easy to see that \scode{min} is a problem: calculating a minimum should not be part of finding a most frequent bigram. Even after the programmer understands the problem, they still need to find a differentiating example that rules out $q_3$. Because the use of \scode{min} in $q_3$ is a minimum over a list of the bigrams in the input, the programmer comes up with an example where the desired bigram is the \emph{largest} one (lexicographically), as in $\sigma_3$.

In this interaction, the programmer had to express the explicit knowledge (``do not use min'')  implicitly through examples. Coming up with examples that avoid \scode{min} requires deep understanding of the program, which is then only leveraged implicitly (through examples).
Even then, there is no guarantee \scode{min} will not recur -- as we will show in \Cref{insufficient}, it is impossible to completely remove it in this model.
%However, in
In this case, since the programmer already knows that they want to avoid programs using \scode{min}, it is beneficial to let them communicate this information explicitly to the synthesizer.
\DONE{SH: can add a forward reference "In fact, as we will see in..., there is \emph{no implicit way} to completely get rid of min in this example." (then in the "key aspects" it will be ok to mention insufficiency). Maybe add this before the "However"  }

%\TODO{can we also say here that it is actually not so easy to see that min is the problem? I want to also motivate the debug information}

\begin{table}[t]
  \centering \small
    \begin{tabular}{ll}
  \multicolumn{2}{l}{\textbf{Task:} find the most frequent bigram in a string} \\
  \hline
%  Step  & PBE \\
%  \hline
  Initial & \multirow{2}{*}{\scode{"abdfibfcfdebdfdebdihgfkjfdebd"}$\mapsto$\scode{"bd"}}\\
  specifications & \\
  \hline
  \begin{tabular}{c}Question \\ $q_1$\end{tabular} & \begin{lstlisting}
input//"abdfibfcfdebdfdebdihgfkjfdebd"
.takeRight(2)//bd\end{lstlisting} \\
\hline
\multicolumn{2}{l}{\textbf{Problem:} \scode{takeRight} will just take the right of a given string}\\
\multicolumn{2}{l}{\textbf{Idea:} \scode{takeRight} will never be useful since we always want to consider}\\
\multicolumn{2}{l}{every element. Remove \scode{takeRight} from the result program.}
\\
  \hline
  Answer & Remove(\scode{takeRight(2)}) \\
  \hline
  \begin{tabular}{c}Question \\ $q_2$\end{tabular} & \begin{lstlisting}
input //"abdfibfcfdebdfdebdihgfkjfdebd"
.drop(1)//"bdfibfcfdebdfdebdihgfkjfdebd"
.take(2) //"bd"
  \end{lstlisting}\\
\hline
\multicolumn{2}{l}{\textbf{Problem:} this program crops a given input at a constant position}\\
\multicolumn{2}{l}{\textbf{Idea:} we don't want to crop anything out, so these functions have no place}\\
\multicolumn{2}{l}{in the result program.}\\
\hline
Answer  & Remove(\scode{drop(1).take(2)}) \\
\hline
\begin{tabular}{c}Question \\ $q_3$\end{tabular}& \begin{lstlisting}
input //"abdfibfcfdebdfdebdihgfkjfdebd"
.zip(input.tail) //List((a,b),(b,d),(d,f),...
.take(2) //List((a,b),(b,d))
.map(p => p._1.toString + p._2) //List("ab","bd")
.max //"bd"
\end{lstlisting}\\
\hline
\multicolumn{2}{l}{\textbf{Problem:} while the beginning of this program is actually good (dividing}\\
\multicolumn{2}{l}{the program into bigrams) and so is the mapping of a $2$-tuple to a string,}\\
\multicolumn{2}{l}{\scode{take(2)} truncates the bigram list.} \\
\multicolumn{2}{l}{\textbf{Idea:} preserve what is good in the program and remove \scode{take(2)} on its}\\
\multicolumn{2}{l}{own and not just as part of a sequence.}\\

\hline
Answer  & \begin{tabular}{l}Affix(\scode{zip(input.tail)})\\Remove(\scode{take(2)})\\Retain(\scode{map(p => p.\_1.toString + p.\_2)})\end{tabular} \\
\hline
%$\vdots$\\
    \end{tabular}%
  %\caption{Handling programs overfitted to the examples with more examples}\label{overview-walkthrough}%
  \caption{Providing granular, syntactic feedback.}\label{gim-walkthrough}
\end{table}%

\subsection{Interaction through a granular interaction model}
GIM improves PBE by employing a richer, \emph{granular} interaction model. On the one hand, the synthesizer supplements the candidate programs by debug information that assists the programmer in understanding the programs, and identifying good and bad parts in them. On the other hand, the user is not restricted to providing semantic input-output examples, but can also mark parts of the program code itself as parts that must or must not appear in any future candidate program.
This allows the user to provide explicit, syntactic, feedback on the program code, which is more expressive, and in some cases allows the synthesizer to more aggressively prune the search space\DONE{I guess this is here for aiding the synthesizer, but it's kind of wtf. SH: It is ok. But "prune" is not specific enough (even removing a single program is pruning). Maybe want to say "aggressively prune" or "more aggressively"}.

The GIM interaction model for the same task of finding the most frequent bigram is demonstrated in \Cref{gim-walkthrough}.
%Consider the same example of finding the most frequent bigram, now with a granular interaction model.
Question 1 is as before: the synthesizer produces the candidate program  \scode{input.takeRight(2)}. In contrast to classical PBE, the granular interaction model provides additional debug information to the user, showing intermediate values of the program on the examples. This is shown as comments next to the lines of the synthesized program. For $q_1$, this is just the input and output values of the initial example. In the next steps this information would be far more valuable.

Given $q_1$, the programmer responds by providing \emph{granular feedback}. Using GIM it is possible to narrow the space of programs using syntactic operations. Presented with \scode{input.takeRight(2)}, the user can \emph{exclude} a sequence of operations from the vocabulary, in this instance \scode{takeRight(2)}, ruling out \emph{any program} where \scode{takeRight(2)} appears. This also significantly reduces the space of candidate programs considered by the synthesizer.

The synthesizer responds with $q_2$. Note that in such cases the debug information assists the programmer in understanding the program, determining whether it is correct, or, as in this case, identifying why it is incorrect.  To rule out $q_2$, the user rules out the sequence \scode{drop(1).take(2)}, as the debug information shows the effect (``take the second and third character of the string''), and the user deems it undesirable at any point in the computation to truncate the string, as all characters should be considered.

The synthesizer responds with $q_3$. This candidate program contains something the programmer would like to preserve: using the debug information, they can see that the prefix \scode{input.zip(input.tail)}  creates all bigrams in the string. The user can mark this prefix to \emph{affix}, or to make sure all candidate programs displayed from now on begin with this prefix. This removes all programs that start with any other function in $\vocab$, effectively slicing the size of the search space by $|\vocab|$. Another option available to the user (multiple operations stemming from the same program are not only allowed but encouraged) is to exclude \scode{take(2)} since the resulting truncation of the list is undesirable.

Eventually, the synthesizer produces the following program:
\begin{lstlisting}
input//"abdfibfcfdebdfdebdihgfkjfdebd"
.zip(input.tail)//List((a,b),(b,d),(d,f),(f,i),(i,b),(b,f),...)
.map(p => p._1.toString + p._2)//List("ab","bd","df","fi","ib",...)
.groupBy(x => x)//Map("bf"->List("bf"),"ib"->List("ib"),...)
.map(kv => kv._1 -> kv._2.length)//Map("bf"->1,"ib"->1,"gf"->1,...)
.maxBy(_._2)//("bd",4)
._1//"bd"
\end{lstlisting}
which does not discard any bigram, counts the number of occurrences, and retrieves the maximum. This program is accepted.

Below we summarize the key aspects of GIM, as demonstrated by the above example.

\paragraph{Key Aspects}

\begin{itemize}
\item Interaction Model: granular interaction in both directions: the synthesizer provides debug information on intermediate states of the program, the programmer can provide feedback (keep/discard) on parts of the program, in addition to input-output examples.
\item Assisting the User: our approach assists the user in two ways. First, the synthesizer supplements candidate programs with debug information that helps the programmer understand the good and bad parts of a candidate program. Second, the ability to give explicit feedback on the code itself provides an alternative (and complementary) way to interact with the system without crafting potentially complicated differentiating examples.
\item Insufficiency of Examples: examples are both inconvenient and insufficient to communicate a programmer's intent. Other operations are needed to allow a programmer to filter programs not only according to semantic equivalence but also according to additional criteria such as readability, best practices and performance.
\end{itemize}

%\input{background}
%\input{prog-enum}
%\section{The Granular Interaction Model}\label{model}

%lower level of control

%syntactic operations

%debugging

%data operations

\section{Background}

%\TODO{SH: I think that we only need a short paragraph on the "Candidate program space" which will briefly cover functional programming, how we denote such programs, and also the vocabulary. We also need to explain somewhere what we mean by "examples". Maybe we can do it in the next section, where we use them and talk about their insufficiency.}

In this work we address synthesis of functional programs. Below we provide the necessary background. % some background on functional programming and synthesis.

%\para{Functional programming}
%Functional programming is a programming paradigm that breaks a desired operation into a composition of functions, and as such also prefers immutable objects. A purely functional solution to a problem would be broken down into a stateless composition of functions $h(g(f(x)))$. Functional programming coexists with object-oriented programming in languages such as Scala, in which case a popular paradigm is a functional composition of object methods, where $f(self)$ is an object method of $f$, which returns a second object (to support immutability) for which $g$ is an object method, allowing the program $g(f(self))$, which returns a third object, and so forth.
%
%notation: f.g.h == h(g(f()))
\para{Notation of functions}
We interchangeably use the mathematical notation $h(g(f(x)))$ for the functional composition called on object $x$ and the Scala notation $x.f.g.h$ (in Scala, a function application with no arguments does not require parentheses).
%Additionally, we use shorthand to denote partially applied functions. \TODO{SH: do we ever use it? can merge this explanation with what we say in the vocabulary paragraph}
%%For instance, if type $\tau$ has a method $foo$ that takes one parameter and we wish to affix the parameter's value, we will define $f=foo(5)$ and then apply $f$ to any objects of type $\tau$.
%For instance if $plus(x,y)$ is a method with two integer parameters, and we wish to affix $y$, we can partially apply the function and save it as $inc(x)=plus(x,1)$ and then apply it to a single numeric argument. Specifically, if the remaining argument is the self reference of an object method, this can be used as a way to create a method that can be applied to any object of the type with no additional arguments.

For a functional program $m$, we denote $\sem{m}$ as the function that the program computes. Formally, $\sem{m}: D\rightarrow D\cup \{\bot\}$ maps every element $i$ in the domain, $D$, either to the element in $D$ that the program outputs on  $i$, or to an error (compilation or runtime) $\bot \not \in D$.

\para{Vocabulary and the candidate program space}
The candidate program space consists of programs %$m$
of the form \scode{input.f$_1$.$\dots$.f$_{n-1}$.f$_n$} (in Scala notation), or $f_n(f_{n-1}(\dots f_1(input)\dots ))$ (in mathematical notation), where each $f_i$ is a method from a predefined vocabulary $\vocab$. Object methods that accept arguments are handled by partially applying them with predefined arguments, such as constants, lambda functions or variables in the context, leaving only the self reference as an argument.
%The space of all programs, in which we expect $m^*$, the target program, to be, is spanned by $\vocab$.
Generally, the candidate program space includes every program in $\vocab^*$, but we notice that for some programs there are compilation errors as not all $f\in \vocab$ is applicable to all objects.
%We define a program to be valid for an input type if no compilation error occurs.

\para{Programming by Example (PBE)}
Programming by Example is a sub-class of program synthesis where all communication with the synthesizer is done using examples. The classic PBE problem is defined as a pair $(\examples,\mathcal{L})$ of initial examples $\examples$ and target language $\mathcal{L}$, where each example in $\examples$ is a pair $(i,o)$ of input $i \in D$ and expected output $o \in D$.
The result of the PBE problem $(\examples,\mathcal{L})$ is a program $m$, which is a valid program in $\mathcal{L}$, that satisfies every example in $\examples$, i.e., $\sem{m}(i) = o$ for every $(i,o)\in \examples$. \ignore{PBE has become widely popular since partial specifications are easier to create than full specifications written in a logic.} Since there might be more than one program $m$ in the language $\mathcal{L}$ that matches all specifications, the iterative PBE problem was introduced. In the iterative model, each candidate program $m_i$ is presented to the user, which may then accept $m_i$ and terminate the run, or answer the synthesizer with additional examples $\examples_i$ which direct it in continuing the search.

\section{The Insufficiency of Examples}\label{insufficient}

In this section, we show the importance of extending the user's answer model beyond input-output examples. We examine in more formal details the scenario described in \secref{overview-pbe}, where the user has seen an undesirable program component and would like to exclude it specifically. We will show that this is not always possible, i.e., that examples are insufficient to communicate the user's intent.
%, and introduce empirical results that show the extent of the problem for the synthesis user.

As seen in \secref{overview-pbe}, the user wishes to rule out the function \scode{min}, but simply providing an example to rule out the current program might not be enough to remove \scode{min} from \emph{all} candidates to ensure it never recurs.
We now formally prove it is \emph{impossible} to completely remove methods like \scode{min} from the search space using examples.
%We have seen in the walkthrough in \secref{overview}, where \scode{min} keeps popping up in new ways despite being given differentiating examples.
%We will now formally prove the impossibility of removing methods like \scode{min} from the search space using examples.

We recall the definition of equivalence between programs. Programs $m_1$ and $m_2$ are \emph{equivalent} if $\sem{m_1} = \sem{m_2}$. %Note that equivalent programs may still differ in other measures: readability, best practices, performance, etc.
Using this we prove the following claim:

\begin{claim}\label{impossibility-claim}
Let $v \in \vocab$ be a letter such that there exists a program $m$ that is equivalent to $m^*$ and contains $v$.
Then examples alone cannot rule out the letter $v\in\vocab$ from candidate programs.
\end{claim}
%\begin{claim}\label{impossibility-claim}
%Let $v \in \vocab$ be a letter in a program displayed by a PBE synthesizer. The synthesis user cannot eliminate $v$ from the search space using examples.
%\end{claim}

%For a functional program $m$, we denote $\sem{m}$ as the function that the program computes. Formally, $\sem{m}: D\rightarrow D\cup \{\bot_C, \bot_R\}$ maps every element $i$ in the domain, $D$, either to the element in $D$ that the program outputs on input  $i$, or to one of two error symbols: $\bot_C, \bot_R \not \in D$. The $\bot_C$ symbol indicates a compilation error when $m$ is not valid for $type(i)$. The $\bot_R$ symbol indicates a runtime error on $i$. (The latter can also be augmented with the type of error, but for the purpose of this paper this is not necessary.)
%\TODO{SH: is it important to talk about the two types of errors here?}

The proof follows since examples can only distinguish between programs that compute different functions.

Next we show that \Cref{impossibility-claim} is applicable to methods that are prevalent in programming languages and extremely useful in some contexts, and therefore are likely to find their way into the vocabularies used in synthesis.
%\TODO{SH: why "inadvertently"? They might really be needed for other tasks. This makes it sounds like we could just eliminate them, but in fact it depends on the task. Need to rephrase. Maybe "are prevalent in programming languages and important for many programming tasks, and therefore are likely to take part in the vocabularies used in synthesis"}
%methods that cannot be removed are prevalent in programming languages and therefore are likely to inadvertently find their way into the vocabularies used in synthesis.
We consider two classes of methods: invertible methods and nullipotent methods.

\para{Invertible methods} are methods for which there exists an inverse method such that applying the two in a pair (in some order) leads back to the initial input.
%that, when applied twice, lead back to the initial input.
For instance, \scode{reverse} on a list or iterator is invertible and its own inverse, as \scode{in.reverse.reverse} will be identical to \scode{in}, i.e. \scode{reverse} is its own inverse method.
%There are also invertible pairs, a pair of methods which cancel each other out, such as
Another example includes \scode{zipWithIndex} and \scode{map(\_.\_1)}, which cancel each other out.
%\TODO{no need for distinction. Invertible means that there is an inverse. It doesn't have to be the same method}
%\HP{If we're not making a distinction I think the best thing to do would be to talk about pairs, it makes the rest simpler}
An invertible method can always be added to the target program along with its inverse, resulting in an equivalent program. Hence, it will never be ruled out by examples.
%Once an invertible method appears an even number of times in the program, or a method pair appears in sequence, it will never be ruled out by examples.

\para{Nullipotent methods} are methods that, when applied, lead to the same result as not being applied. While this is often context-sensitive, e.g. calling \scode{toList} on a list or \scode{mkString} on a string, there are calls that will always be nullipotent, such as \scode{takeWhile(true)}. Because some methods are nullipotent only in a certain context, they may be in a synthesizer's vocabulary, and end up in the program space in contexts where they are nullipotent. It is easy to construct a program that contains nullipotent methods and is equivalent to the target program. Hence, similarly to invertible methods, these methods cannot be eliminated by examples.

%\Cref{impossibility-claim} is also applicable to methods such as \scode{min} that do not fall into these categories:

\begin{sloppypar}
Returning to our example, for the target program in \secref{overview}, \scode{input.zip(input.tail).map(p => p.\_1.toString + p.\_2).groupBy(x => x).map(kv => kv.\_1 -> kv.\_2.length).maxBy(\_.\_2).\_1}, let us now construct an equivalent program
%to the target program
by appending to it an invertible pair of functions in sequence: \scode{sliding(2).min}. The function \scode{sliding(2)}, when applied to a string of length $2$ will return \scode{List("dc")}, and \scode{min} when applied to list of size $1$ will return the only member of the list. This means there is a program that is equivalent to $m^*$ on every input, and contains \scode{min}. As such, given any number of examples applied \scode{min}, a letter from $\vocab$, will not be ruled out entirely.\end{sloppypar}

This construction is possible for many target programs, showing that it is often impossible to discard an undesirable member of the alphabet or an undesired sequence using examples alone.

%Not only is the user unable to rule out such undesired components, but in some cases these methods overshadow desired components.
%Since these methods are common, we can see that applying \Cref{impossibility-claim} by constructing a program that is equivalent to the target program and contains a component that cannot be ruled out is often very easy.
Furthermore, since many existing PBE synthesizers prune very aggressively based on \emph{observational equivalence}, or equivalence based only on the given examples, programs that do not include the undesired component may not be available anymore as they've been removed from the space.

These properties leave us with the need to define a more expressive, more granular model.

%\begin{corollary}
%A set of all programs with the letter $v\in\vocab$ cannot be ruled out if the equivalence class of $m^*$ contains a program with $v$.
%\end{corollary}

The practical implications of \cref{impossibility-claim} are discussed in \cref{study-results}, which examines the existence of method sequences deemed undesirable by users in candidate programs. The data as well as opinions collected from users show that the inability to remove an undesirable letter from the alphabet has real-world consequences, which affect the user's frustration with the synthesizer (see \cref{excluded-table}).

\section{The Granular Interaction Model}\seclabel{granular}

In this section, we describe the Granular Interaction Model (GIM) mechanism, which extends the PBE model with additional predicates. Namely, predicates in GIM include examples, but also additional predicates. The key idea is to add a broader form of feedback from the user to the synthesizer than has been available in PBE. We begin by describing the operations and the type of feedback that each one of them allows the user to provide the synthesizer with, and discuss the observed uses of each.

\subsection{Granular predicates}\label{predicates-def}

In the setting of functional compositions, we choose to present GIM with three syntactic predicates. We refer to these predicates as granular since they impose constraints on \emph{parts} of the program, rather than on its full behavior, as captured by the function it computes or its input and output types. We will also discuss other, possible predicates.

Given a candidate program $m=f_n(f_{n-1}(\dots input \dots ))$ we introduce the following predicates, to be tested against other programs $m'=f'_m(f'_{m-1}(\dots input \dots ))$
\begin{enumerate}
  \item $\exclude(f_i,\dots,f_j)$ where $i\leq j$: will hold only for programs $m'$ where $\neg\exists k. f'_k = f_i \wedge \dots \wedge f'_{k+i-j}=f_j$
  \item $\retain(f_i,\dots, f_j)$ where $i\leq j$: will hold only for programs $m'$ where $\exists k. f'_k = f_i \wedge \dots \wedge f'_{k+i-j}=f_j$
  \item $\affix(f_0,\dots,f_i)$: will hold only for programs $m'$ where $\forall j\leq i.f_j=f'_j$.
\end{enumerate}

The $\exclude$ operation rules out a sequence of one or more method calls as undesirable. For the example in \secref{overview}, to rule out \scode{min} the user would simply add the predicate $\exclude(\scode{min})$. However, should the user rule out a sequence longer than a single method, this would apply to the sequence as a whole: the predicate $\exclude(\scode{reverse},\scode{reverse})$ does not exclude the \scode{reverse} method, only two consecutive invocations of it that cancel out.

The $\retain$ operation defines a sequence that must appear in the target program. It is similarly defined for sequences: when applied to a single method, it forces the method, and when applied to a sequence it forces the sequence, in-order. It can be seen, essentially, as creating a procedure and then deeming that procedure as desirable.

However, since the $\retain$ is not dependent on the location of the procedure in the program, we add an additional predicate for not only setting a procedure, but forcing its location to the beginning of the program. The $\affix$ predicate will essentially narrow the search space to sub-programs that come after the desired prefix.

\para{Additional predicates}
%This section has focused on three syntactic operations as an extension to the interaction model. However, they are not alone - 
As these three operations are highly expressive and easy to understand, we have centered the experiments in this paper around them, but they are, by no means, the only possible predicates.
Many other granular operations exist. For instance, the user can reason about intermediate states of the program by demanding or excluding certain intermediate states for a given input. A user can also require an error, or an error of a certain kind, for a given input. \Cref{prog-selection} will expand on the reasons to select certain expansions to the interaction model over others.

\subsection{Adding a debugging view of the code}\seclabel{debugview}

GIM assumes an interaction with users that are comfortable with reading code. This means not only that more can be expected from them, but that they can be assisted in ways not generally offered by a regular synthesizer. In the same way the interaction from the user to the synthesizer can be granulated, so can the interaction from the synthesizer to the user. %In its simplest instance, the synthesizer can present the user with intermediate information of the candidate program.

PBE tools like FlashFill only show the user the output of running the program on an input. Other tools that do show code show the program while guaranteeing it satisfies all examples in $\examples$. In a functional concatenation, it is possible to show the user the result of each subprogram, on each $e\in\examples$. This means that even if some $f\in\vocab$ is not familiar to the user, they can still gauge its effect and determine whether or not that effect is desired \emph{by example}.% rather than having to look up every unfamiliar method.

\begin{Example}
Let us consider the case where input is a list of strings, and the user is presented with the candidate program \scode{input.sliding(3).map(l => l.mkString)}. While they are familiar with the \scode{mkString} method, which formats a list into a string, and with mapping a list, they have never encountered \scode{sliding}.

The user could look up the method and read up on its behavior. However, oftentimes its behavior will be simple enough to understand by its operation within the program. Therefore, if the user is provided with the intermediate states of the program like so:
\begin{lstlisting}
input //List("aa","bb","cc","dd","ee")
.sliding(3)//List(List("aa", "bb", "cc"),List("bb", "cc", "dd"),...)
.map(s => s.mkString) //List("aabbcc","bbccdd","ccddee")
\end{lstlisting}
they can understand that \scode{sliding} returns a list of sublists of length $n$ beginning at each position in the list -- a sliding window of size $n$.
\end{Example}

\subsection{Enabling the User}

After we have introduced the formal framework for predicates, we now wish to leverage it to create a user interaction model. We suggest the following iterative process, which we have implemented for the user study in \cref{userstudy}.

\begin{figure}
%\centering
\begin{tabularx}{\columnwidth}{c}
%\captionof{figure}{Program display with debug information}\figlabel{ui1}
%\begin{minipage}[t]{.5\textwidth}
%\centering
\includegraphics[width=0.98\columnwidth]{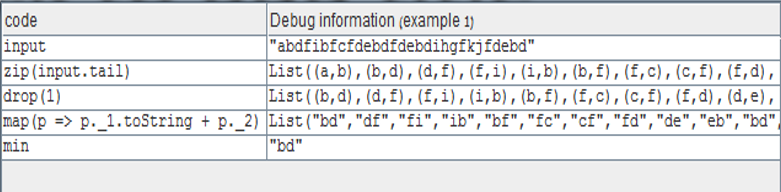}
%\end{minipage}%\hfill%

\\
%\captionof{figure}{User selecting a sequence for exclusion}\figlabel{ui2}
%\begin{minipage}[t]{.5\textwidth}
%  \centering
\includegraphics[width=0.98\columnwidth]{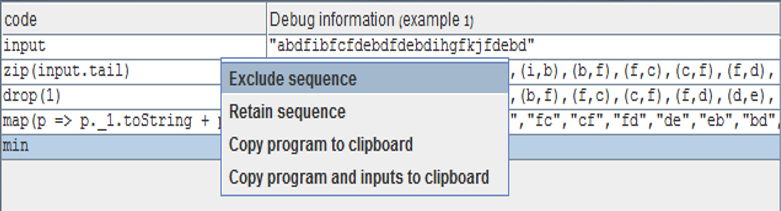}
%\end{minipage}
\end{tabularx}
\caption{Program with debug information, a sequence selected for removal}\label{ui1}
\end{figure}

A candidate program is displayed to the user alongside the debug information. The top image in \cref{ui1} shows this in our UI. The user is now able to study the program and accept or reject it.

If the user is dissatisfied with the program, and would like to reject it, the goal is to allow them to directly express the source of dissatisfaction as predicates as easily as possible. Towards this end, we let the user point out a portion of the program (e.g. by right-clicking it) and mark it as desirable or undesirable, as seen in the bottom image in \cref{ui1}.

This process of easily providing feedback on the  program turns predicates into a convenient tool for feedback to the synthesizer.

%\section{Program selection}\label{prog-selection}
\subsection{Enabling the synthesizer}\label{prog-selection}
%TODO hijacking this section to talk about the impl of $Select$.

As we have seen, the choice of predicates is crucial from the user's perspective. However, it is also important for the synthesizer to be able to use them in maintaining and updating a representation of the search space. To complete this section, we show how the predicates described in \cref{predicates-def} are naturally utilized by a synthesizer for the domain of linear functional concatenations.

%The process of selecting the next program from the current state of the synthesizer $A$ has been denoted $Select(A)$. This process will vary from synthesizer to synthesizer, depending on both the method of synthesis used and the types of predicates allowed. Here we demonstrate this notion with a naive enumerating synthesizer which will be used in \secref{experiments} and a probabilistic synthesizer.
%\TODO{we also show that the choice of predicates fits well with these two synthesizers. Namely, they can benefit from the predicates for pruning their search space}

\para{Enumerating synthesizer}
The state of the art in program synthesis hinges on enumerating the program space in a bottom-up fashion~\cite{alur2017scaling,feser2015synthesizing,alur2016sygus}.
For the domain considered in this paper, bottom-up enumeration consists of concatenating method calls to prefixes already enumerated, starting with the program of length $0$, \scode{input}. This enumeration is restricted by types, i.e. by compilation.
%We demonstrate the usefulness of GIM using a naive enumerating synthesizer will essentially enumerate the set $\vocab^*$ of sequences over $\vocab$ by length: first the program of length $0$ (\scode{input}), next all programs of length $1$ that compile with regards to $type(input)$, and so on.
The search space in this synthesizer can be represented as an edge-labeled tree where the root is the program $input$ and each edge is labeled by a method name from $\vocab$.
Each finite-length path in the tree represents the program that is the concatenation of every label along the path.
The tree is initially pruned by compilation errors (i.e., if $f \in \vocab$ does not exist for the return type of $m$, it will be pruned from the children of the node representing $m$). It now represents the candidate program space for %a synthesizer state %$\top$.
an unconstrained synthesizer state.

%In such a synthesizer,
%we notice that the predicates are divided between two phases, one affecting the internal representation of the state, and one affecting the selection of the program being displayed: in a tree representation
We can see that every program deemed undesirable by the operations $\affix$ and $\exclude$ cannot be extended into a desirable program. Therefore these extensions can be discarded and the tree representing the candidate space can be pruned at the nodes of these programs.

This is an example of predicates that are \emph{well suited} to the representation of the state of the synthesizer, in that they not only aid the user but also help guide the search of the space.
Since the combination of the enumeration and these predicates is monotone, a program that was pruned from the search space will never need to be looked at in a future, more constrained iteration. This means that the synthesizer does not need to be restarted across iterations -- however, even if it is, these predicates will allow it to construct a much smaller search space to begin with.

\section{Evaluation}\seclabel{experiments}

To evaluate our approach, we compared three interaction models:
\begin{enumerate}
  \item PBE: replicating the state of the art in synthesis, the user can communicate with the synthesizer via input-output pairs.
  \item Syntax: testing the new operation set proposed in \secref{granular}, the user can communicate with the synthesizer via syntactic predicates on the program.
  \item GIM: testing the full model, the user can communicate via both sets of predicates.
\end{enumerate}
We limit the test of the granular interaction model to three operations that are relevant to functional compositions and are easy to understand. Therefore, we select the operations detailed in \cref{predicates-def} as our basic set of granular operations.

We have conducted two studies:
\begin{compactenum}
\item A study of ideal sessions with different operations (i.e., families of predicates) for a set of benchmarks.
%\item A study of the size of the search space and its ambiguity, and of the effect of different operations (i.e., families of predicates) on the size of the space.
%\TODO{SH: the actual experiment shows the convergence rate. It is not exactly the effect on the size of the space. Perhaps rephrase}
 %left given the examples in $14$ programming tasks.
 %This study also considers the ideal synthesis sessions for these tasks with several families of predicates.
\item A controlled user study which tests the usability of a GIM synthesizer for  programmers and the benefits when measured against a control group using PBE.
\end{compactenum}

%We limit the test of the granular interaction model to three operations that are relevant to functional compositions and are easy to understand. Therefore, we select the operations detailed in \cref{predicates-def} as our basic set of granular operations.
% and add them to examples to study the usefulness of GIM.

\para{Synthesizer}
We implemented a simple enumerating synthesizer  described in \cref{prog-selection} in Scala, using the nsc interpreter (used to implement the Scala REPL). The vocabulary $\vocab$ is provided to the algorithm, and programs are compiled and evaluated on the inputs.

In order to support the study in \cref{sspace}, the synthesizer accepts input of additional examples, rejection of the current program, or of $\affix$, $\exclude$ and $\retain$ predicates. In order to support the user study in \cref{userstudy}, it also precomputes the space of valid programs.

\subsection{Problem set}\label{benchmarks}

We performed the studies using a set of functional programming exercises %that we collected
from three different domains: strings, lists and streams. The exercises were collected from Scala tutorial sites and examples for using MapReduce. The tasks, described in \tabref{questions}, were each paired with a vocabulary and an initial set of examples.

\para{Discussion}
As seen in \tabref{questions}, the set of valid programs is significantly smaller than $|\vocab|^{|m^*|}$, but in many cases the space still contains thousands or tens of thousands of programs.
There is also fair amount of inherent ambiguity over the initial example set $\examples_{init}$, as can be seen in the ``reject only'' column, representing the set of all programs up to length $|m^{*}|$ that match $\examples_{init}$. This means that, even limiting the search space to the known length of the target program we would start with %an observational equivalence class ($\equiv_{\examples_{init}}$) of hundreds or thousands of programs that need to be filtered by the user.
hundreds or thousands of matching programs that need to be filtered by the user.

%\subsection{Search spaces and operations needed}\label{sspace}
\subsection{Ideal synthesis sessions}\label{sspace}

% Table generated by Excel2LaTeX from sheet 'Sheet3'
\begin{table*}[t]
\footnotesize
  \centering
    \begin{tabular}{r|ll|cccccccc}
    &          &  &            &                      &         &            & \multicolumn{4}{c}{Number of Candidates} \\ \cline{8-11}
    &          &  &            &                      &         & candidate  & reject &            &  &  \\
    &Benchmark &  & $|\vocab|$ & $|\examples_{init}|$ & $|m^*|$ & space size & only   & PBE        &Syntax     &  GIM \\
    \hline
    \multirow{5}[0]{*}{\begin{sideways}strings\end{sideways}} & dropnthletter & Drop every 5th letter in a string & 20    & 1     & 3     & 280 & 4     & 2     & 3     & 2 (1) \\
          & freqbigram & Most frequent bigram in a string & 19    & 1     & 6     & 118261 & 674   & 4     & 8     & 6 (0)\\
          & frequword & Most frequent word in a string & 25    & 1     & 4    & 4853 & 126   & 5     & 8     & 6 (1)\\
          & linesinfile & Number of lines in file & 20    & 1     & 2    & 47 & 4     & 3     & 4     & 3 (1)\\
          & nonemptylines & Number of non-empty lines in file & 21  & 1    & 3 & 1664 &  29   &  2   &   3  &  3 (1)\\
          \hline
    \multirow{6}[0]{*}{\begin{sideways}lists\end{sideways}} & anagrams & Group words that are anagrams & 17    & 1     & 6     & 13554 & 12    & 3     & 3     & 3 (0) \\
          & histogram & Create a histogram of number list & 12    & 1     & 5    & 4208 & 37    & 3     & 9     & 3 (1)\\
          & median & Find the median of a list of numbers & 20    & 1     & 4     & 71211 & 1663  & 6     & 14    & 9 (1)\\
          & posinlist & Get all positive numbers from list & 20    & 1     & 2   & 190 & 17    & 3     & 4     & 4 (1)\\
          & sudokusquare & Validate a square in sudoku & 17    & 1     & 5    & 1602 & 118   & 3     & 7     & 4 (0)\\
          & sumsquares & Sum of squares of a list of numbers & 20    & 1     & 2    & 120 & 2     & 2     & 2     & 2 (0)\\
          \hline
    \multirow{3}[0]{*}{\begin{sideways}streams\end{sideways}} & bitstream & Next integer from a stream of bits & 17    & 2     & 4    & 3717 & 101   & 2     & 8     & 5 (1)\\
          & numhashtags & Count hashtags in a stream of tweets & 15    & 1     & 7    & 11527 & 2     & 2     & 2     & 2 (0)\\
          & slidingavg & Average next three values from every index & 25    & 1     & 4    & 60479 & 125   & 2     & 2     & 2 (0)\\

    \end{tabular}%
  \caption{
  %The size of the starting conditions (vocabulary and initial examples), size of the target program $m^*$, and size of the candidate program space (up to programs of length $|m^*|$ for each task.
  %In addition, the number of questions (candidate programs) asked by the synthesizer for an optimized ideal run of the synthesizer under four sets of available predicates. This includes the target program, which is the final candidate. The number in parentheses next to the number of ``all'' candidates indicates how many candidates were responded to with an example, when given the option to select from both examples and syntactic operations.
  The test setup of $14$ synthesis experiments, showing the ambiguity inherent in $\examples_{init}$, and the number of iterations to the target program in an ideal synthesis session with each available set of operations. Parentheses indicate examples used.
 }\tablabel{questions}
\end{table*}%

\para{Experimental questions} For each task in the problem set we answer the following question: under the ideal conditions of an expert user and knowledge of the target program, how many questions (i.e. candidate programs) are posed to the user for each predicate family?

\para{Test setup} In order to test these questions, each task in the problem set was run in four settings:
%optimize the course of each session
%set of predicates to create a run with fewer iterations and fewer predicates provided.
\begin{compactitem}
\item \textbf{Reject only}: no operations other than rejecting the current example. This is an enumeration of the programs that match the initial example set. %This gives a measure of the inherent ambiguity in the initial example set, through the count of how many valid, matching programs will need to be eliminated.
%    The results of this run also allowed us to thoroughly explore different strategies of additional example selection for the PBE experiments.
\item \textbf{PBE}, \textbf{Syntax}, and \textbf{GIM}: as described above, all with the addition of a reject operation.
\end{compactitem}
Examples and other predicates were selected by an expert user (author of this paper) making an effort to create a run with fewer iterations and more aggressive pruning of the space in each iteration.

\para{Results}
\tabref{questions} shows the results for each of the programming tasks. As can be seen from the table,
in ideal (i.e. thoroughly optimized, expert user) runs, the number of questions produced by the synthesizer for a PBE run was lowest. This was not unexpected: carefully selected examples are a fast way to differentiate between programs. The subject of examples selected in less ideal conditions is left to the following section. But we also see that in a run allowing all predicates the number of questions asked was lowered substantially compared to syntax only, without involving more than one example.
%\end{compactitem}

\subsection{User study}\label{userstudy}

To test the interaction between programmer and synthesizer, we conducted a user study, where we compared the interaction of programmers with the synthesizer using the three families of operations: %three sets of available operations:
%\TODO{SH: very similar to the operations in the previous subsection. What is the difference? Can we use the same names and unify? No need to repeat}
\textbf{PBE} (control),
\textbf{Syntax}, and
\textbf{GIM}.

%\begin{inparaenum}[(i)]
%\item PBE (control): consisting of input-output examples,
%\item Syntax: consisting of the new operation set proposed in \secref{granular}, and
%\item GIM: the full model which includes all above operations.
%\end{inparaenum}

\para{Research questions}
We examine the following questions:

\begin{compactenum}[(1)]
  \item \emph{Are answers consisting of syntactic predicates easier or faster to generate than example predicates?}\label{faster}
  This question is examined in two different ways: first, for each task the average and median iteration times with the synthesizer are compared between the control group (PBE) and the Syntax group.
  Second, when users are allowed both (GIM), the time spent on iterations where they provided examples is measured against their average time.%, to see if there is an increase.

  \item \emph{Is the total time to solution improved by adding or exchanging the available predicates?}\label{total-time}

  \item \emph{Are users able to reach a correct program using each of the predicate sets?}\label{correctness-q}

  \item \emph{Do users prefer examples?}\label{preference}
  This question examines the choices made by the participants in the GIM group, which had a choice between all possible predicates. We test how often examples were chosen, and whether the task being solved had an effect on this preference.

  \item \emph{Are users in PBE sessions distracted by undesirable sequences that cannot be removed?}\label{unremovable}
  We check the recurrence in the PBE group of sequences that were deemed undesirable by users in the Syntax and GIM groups, and try to determine whether these repeat enough to distract the user.
  We also check for the acceptance of equivalent programs with superfluous elements as mentioned in \cref{impossibility-claim}.
  In addition, we bring some anecdotal opinions volunteered by participants.
\end{compactenum}

\vspace{0.1in}

Most questions are examined on all participants. We show data for the small set of users experienced in Scala against those new to Scala when the difference is of interest.

\para{Test setup}
$32$ developers participated in the study. They consist of $7$ undergraduates in their final year of a CS degree, $9$ graduate students in CS, most with a history as developers outside academia, and $16$ industry developers employed by four different companies. Of the $32$, $8$ had prior experience with the Scala programming language.

%Our study included three test groups,
The participants in the study were evenly distributed between three test groups: PBE, Syntax and GIM.
%family of predicates:
%\begin{enumerate}
%  \item PBE (control): replicating the state of the art in synthesis, the participant was able to communicate with the synthesizer via input-output pairs.
%  \item Syntax: testing the new operation set proposed in \secref{granular}, the participant was able to communicate with the synthesizer via syntactic predicates on the program.
%  \item GIM: testing the full model, the participant was able to communicate via both sets of predicates.
%\end{enumerate}
%
%\TODO{to test setup}
%The participants were evenly and randomly distributed between groups.
Each participant was randomly assigned to one of the test groups. Not all participants performed all tasks (scheduling constraints were cited for the most part). The order of the tasks was randomized for each user.%, there is still a fairly even distribution of sessions (detailed in \cref{all-sumup}) for each task.

%Each participant in the study was randomly assigned to one of three test groups:

The reject operation was not allowed in any group, forcing users %so users were forced % in order to force the users
to provide the process with new information as they would in any state of the art synthesizer, rather than just iterate the program space.

Each participant was asked to use the synthesizer to solve three programming questions. % within the tools available to them.
The three problems---frequword, nonemptylines, and histogram---were selected from the tasks tested in \cref{sspace} because of their high level of ambiguity based on the initial example, and not requiring any additional libraries or definitions outside the Scala standard library to solve (i.e., the programs could be run in a Scala console with no imports or definitions).

Participants were given a short introduction to Scala, if they were not already familiar with it, and aided themselves with a Scala REPL, but no online sources or documentation.

\para{Implementation}
Participants performed the tasks using the UI shown in \cref{ui1}. The space of programs was precomputed by the enumerating synthesizer detailed in \cref{benchmarks} and over the same initial inputs, up to a program length of $6$.
In each iteration a program that uphold every predicate given by the user was selected from the set of programs. Selection used a hash-based criterion to prevent lexicographical ordering and favoring of short programs, in order to also show the user complex programs. At the end of every iteration the user's answers are added to the synthesizer's state and the programs are filtered accordingly. If the precomputed set is exhausted, the user is given the option of starting over or abandoning the current task.

\subsection{User study results}\label{study-results}
We address each question individually.

% Table generated by Excel2LaTeX from sheet 'all-sumup'
\begin{table}[t]
  \centering
  \footnotesize
    \begin{tabularx}{\columnwidth}{@{}
    >{\hsize=23.5pt}X|
    >{\hsize=17pt}X|
    @{\extracolsep{2pt}}c
    @{\extracolsep{5pt}}c
    @{\extracolsep{7pt}}c
    @{\extracolsep{5pt}}c
    @{\extracolsep{7pt}}c|
    @{\extracolsep{1pt}}
    c@{\extracolsep{1pt}}
    c@{\extracolsep{1pt}}
    c}
                          &                   &            & \multicolumn{2}{c}{iteration}  &\multicolumn{2}{c|}{number of}  & & \multicolumn{2}{c}{correct} \\
                          \cline{9-10}
                          &                       & no. of          & \multicolumn{2}{c}{time (sec)} &\multicolumn{2}{c|}{iterations} &    & target  & equiv. \\
                      \cline{4-5} \cline{6-7}
    task                  & group            & sessions & avg & med  & avg & med & finished & answer & answer \\
    \hline
    \multirow{3}[0]{*}{histogram} & PBE                   & 11                    & 163.34                & 131.97                & 2.45                  & 2.0                  & 11                    & 1                     & 10 \\
                          & Syntax           & 9                     & 86.27                 & 59.97                 & 12.11                 & 8.0                  & 9                     & 3                     & 4 \\
                          & GIM                  & 10                    & 98.78                 & 96.18                 & 8.90                  & 7.5                  & 10                    & 5                     & 4 \\
    \hline
    \multirow{3}[0]{*}{\shortstack{no. lines\\ with text}} & PBE                   & 11                    & 170.13                & 168.17                & 2.73                  & 2.0                  & 11                    & 0                     & 8 \\
                          & Syntax           & 8                     & 82.16                 & 60.56                 & 10.50                 & 9.5                  & 7                     & 3                     & 1 \\
                          & GIM                  & 11                    & 78.26                 & 65.78                 & 8.82                  & 8.0                  & 9                     & 4                     & 3 \\
    \hline
    \multirow{3}[0]{*}{\shortstack{most\\ frequent\\ word}} & PBE                   & 11                    & 114.52                & 71.27                 & 4.45                  & 4.0                  & 10                    & 9                     & 0 \\
                          & Syntax           & 10                    & 58.28                 & 50.34                 & 22.10                 & 15.0                 & 8                     & 7                     & 0 \\
                          & GIM                  & 11                    & 79.87                 & 53.84                 & 8.82                  & 8.0                  & 10                    & 8                     & 0 \\
    \end{tabularx}%
  \caption{Summary of the three tasks performed in the user study (all users). %Iteration times and number of iterations (average and median) are shown. The last three columns show correctness figures for each test group: how many sessions ``finished'', or ended in accepting a program, how many of those accepted the target program, and how many accepted an equivalent program containing noop elements.}\label{all-sumup}%
  }\label{all-sumup}%
\end{table}%

\para{Question \ref{faster}:}
The average and median times per iteration are shown in \cref{all-sumup}. Medians are also shown in~\figref{meditertime}.

We examined the distributions of data using the Mann-Whitney test. The threshold for statistical significance selected was $p<0.05$. A significant difference was found in the time per iteration between the control (PBE) group and the syntax-only group for all tests: histogram (131.97s, 59.97s, $\pvaleq{0.03}$), nonemptylines (168.17s, 60.56s, $\pvaleq{0.03}$) and frequword (71.27s, 50.34s, $\pvaleq{0.04}$). A significant difference was found between the control group and the GIM group for two of the three tests: histogram (131.97s, 96.18s, $\pvaleq{0.03}$), nonemptylines (168.17s, 65.78 s, $\pvaleq{0.03}$), but not for frequword (71.27s, 53.84s, $\pvaleq{0.058}$). Additionally, a significant difference was found between the syntax-only group and the GIM group for one test: histogram (59.97s, 96.18s, $\pvaleq{0.047}$), but not for nonemptylines (60.56s, 65.78s, $\pvaleq{0.33}$) or frequword (50.34s, 53.84s,$\pvaleq{0.19}$).

These results imply that, with the exception of the frequword test for the GIM group, using either a syntax-only or both syntax and example predicates, there is a speedup in iteration time from solving the same problem in PBE alone. Additionally, with the exception of the histogram task, the slowdown in iteration time between syntax-only and GIM seems to be coincidental.

%We also find statistical significance in the growth in number of iterations from PBE to the other groups in all three tests individually, and no statistical significance in the difference in number of iterations between (except frequword)

In addition, we looked only at the session for users in the GIM group and within each session examined the time to create an example against the average iteration time. There is a slowdown of $19.5\%$ in iteration time with an example, and we see that this difference is statistically significant (75.03s, 90.11s, $\pvaleq{0.049}$).

We can therefore answer question \ref{faster} in the affirmative on both counts: syntactic predicates are faster to generate than examples, both when examining the test groups against the PBE group, and when examining the users with access to both against themselves.

\para{Question \ref{total-time}:}
We noticed a change in the median total time between the control (PBE) and the other groups (Syntax and GIM) indicating a possible slowdown. However, for none of the individual tests, as well as for a unification of all tests, was this change statistically significant ($p > 0.25$ for all). Therefore, while we do not answer question \ref{total-time} in the affirmative -- as the total time was not improved in either of the test groups -- we can also say that the evidence of a slowdown may be coincidental.

\para{Question \ref{correctness-q}:}
%number of participants that were able to reach code that would answer the task given to them, and the number of participants that were able to reach the target program itself rather than stop at an equivalent program
The correctness results in \cref{all-sumup} are visualized in~\figref{correctness}. Aside from the histogram task, completed by all users, all other tasks had some users stopping without accepting a program. The success percentage in reaching any functionally-correct response, is highest for PBE ($100\%, 73\%, 90\%$), lowest for Syntax ($78\%, 57\%, 87\%$), and rebounds with GIM ($90\%,77\%,80\%$) to levels close to the control, even overtaking it for the nonemptylines task.

% Table generated by Excel2LaTeX from sheet 'all-sumup'
\begin{table}[t]
  \centering
  \footnotesize
    \begin{tabular}{l|l@{\extracolsep{2pt}}c@{\extracolsep{2pt}}cr@{\extracolsep{3pt}}r@{\extracolsep{3pt}}r@{\extracolsep{3pt}}r}
                      &                       &                    & sessions & & & & \\
                      &                       & no. of             & used         & \multicolumn{4}{c}{percent examples per user}  \\
                      \cline{5-8}
                      & task                  & sessions           & examples     & avg                   & med              & min         & max \\
    \hline
      \multirow{3}[0]{*}{\shortstack{all\\ users}}                 & histogram             & 10                    & 9                     & 37.6\%                & 35.0\%                & 0.0\%                 & 85.7\% \\
                      & nonemptylines   & 11                    & 8                     & 29.7\%                & 31.0\%                & 0.0\%                 & 66.7\% \\
                      & frequword       & 11                    & 10                    & 36.1\%                & 37.5\%                & 0.0\%                 & 85.7\% \\
      \hline
    users             & histogram       & 2                     & 2                     & 74.1\%                & 74.1\%                & 62.5\%                & 85.7\% \\
    familiar          & nonemptylines   & 2                     & 2                     & 46.7\%                & 46.7\%                & 33.3\%                & 60.0\% \\
    with Scala        & frequword       & 2                     & 2                     & 29.9\%                & 29.9\%                & 22.2\%                & 37.5\% \\
    \hline
    users             & histogram       & 8                     & 7                     & 28.5\%                & 17.7\%                & 0.0\%                 & 66.7\% \\
    unfamiliar        & nonemptylines   & 9                     & 6                     & 25.9\%                & 30.0\%                & 0.0\%                 & 66.7\% \\
    with Scala        & frequword       & 9                     & 8                     & 37.5\%                & 37.5\%                & 0.0\%                 & 85.7\% \\
    \end{tabular}%
  \caption{Proportional part (\%) of examples in the predicates provided by GIM group users. Some used no examples at all, none used only examples.}\label{usedexamples}
\end{table}%

\para{Question \ref{preference}:}
%How often were examples chosen? Did the task being solved have an effect on this preference?
A summary of how often users chose examples appears in \cref{usedexamples} and \cref{Fi:examples-minmaxmed}. We can see a distinction between users familiar with Scala and users who are not. While users familiar with Scala used examples in every task, for users unfamiliar with Scala every task included at least one user avoiding examples-- sometimes as many as $\nicefrac{1}{3}$ of the users. The proportional part of examples out of the total predicates used in the task is fairly low for the entire test group, ranging from $31\%$ to $37.5\%$ (median).

Looking at the data by familiarity with Scala we see that the preference for examples is inverse between the two groups: users familiar with Scala preferred more examples overwhelmingly (over $60\%$ examples for both users) for the histogram task and preferred other predicates for the frequword (most frequent word) task. Conversely, users unfamiliar with Scala preferred examples (but not as overwhelmingly, half the participants over $30\%$) for the frequword task and favored other predicates (half the participants under $20\%$ examples) when solving the histogram task. This seems to suggest a relationship with the difficulty of the task -- histogram is a less trivial problem than frequword. Despite this, even in cases where examples were favored, they were not the only tool used.

% Table generated by Excel2LaTeX from sheet 'Sheet2'
\begin{table}[t]
  \footnotesize
  \centering
    \begin{tabularx}{\columnwidth}{@{}
        >{\hsize=1.5pt}X|
        l@{\extracolsep{2pt}}
        l
        @{\extracolsep{3pt}}>{\hsize=2pt}c
        >{\hsize=2pt}c
        @{\extracolsep{1pt}}c
        @{\extracolsep{2pt}}>{\hsize=3pt}l
    }
    & &  & \multicolumn{4}{c@{\hspace{-1pt}}}{PBE} \\
            \cline{4-7}
                          &                       & GIM/Syntax & \multicolumn{2}{l@{\hspace{-1pt}\vspace{1pt}}}{times seen} & distracting  & \\
                          &                       & users saw & \multicolumn{2}{l@{\hspace{-1pt}}}{in session} & occurrences  & users \\
                                                                \cline{4-5}
                          & removed sequence      & and removed      & min & max                & (average)    & distracted \\
    \hline

    \multirow{5}[2]{*}{\begin{sideways}num of lines\end{sideways}} & tail                  & 84.2\% (16)           & 1                     & 4                     & 2.8                   & 45.5\% (5) \\
                          & takeWhile(c => c != "\textbackslash{}n") & 73.7\% (14)           & 1                     & 4                     & 2.7                   & 54.5\% (6) \\
                          & filterNot(c => c=='\textbackslash{}r' || c=='\textbackslash{}n') & 57.9\% (11)           & 0                     & 3                     & 2.3                   & 27.3\% (3) \\
                          & filter(!\_.isEmpty)   & 27.3\% (3)            & 0                     & 3                     & 2.3                   & 27.3\% (3) \\
                          & tail.takeWhile(c => c != "\textbackslash{}n") & 15.8\% (3)            & 1                     & 4                     & 2.8                   & 45.5\% (5) \\
\hline   \multirow{9}[2]{*}{\begin{sideways}most frequent word\end{sideways}} & takeRight(1)          & 100.0\% (12)          & 0                     & 1                     & 0                     & 0.0\% (0) \\
                          & drop(10)              & 84.6\% (11)           & 0                     & 1                     & 0                     & 0.0\% (0) \\
                          & drop(1)               & 76.5\% (13)           & 0                     & 3                     & 2.5                   & 16.7\% (2) \\
                          & takeRight(6)          & 76.2\% (16)           & 1                     & 3                     & 2.4                   & 58.3\% (7) \\
                          & dropRight(1)          & 71.4\% (15)           & 0                     & 7                     & 3.3                   & 33.3\% (4) \\
                          & take(5)               & 57.1\% (12)           & 1                     & 5                     & 2.8                   & 66.7\% (8) \\
                          & last                  & 42.9\% (6)            & 0                     & 3                     & 3                     & 16.7\% (2) \\
                          & drop(10).drop(1)      & 41.7\% (5)            & 0                     & 1                     & 0                     & 0.0\% (0) \\
                          & takeRight(6).takeRight(6) & 38.1\% (8)            & 1                     & 2                     & 2                     & 8.3\% (1) \\
\hline \hline   \multirow{4}[1]{*}{\begin{sideways}histogram\end{sideways}} & toMap                 & 57.9\% (11)           & 1                     & 1                     & 0                     & 0.0\% (0) \\
                          & map(\_.\_1 -> 1)      & 42.1\% (8)            & 1                     & 1                     & 0                     & 0.0\% (0) \\
                          & zipWithIndex          & 26.3\% (5)            & 1                     & 1                     & 0                     & 0.0\% (0) \\
                          & map(\_.\_1.toInt)     & 15.8\% (3)            & 1                     & 1                     & 0                     & 0.0\% (0) \\

        \end{tabularx}%
  \caption{Frequently-removed method sequences in the Syntax and GIM groups and their occurrence in the PBE group.}
  \label{excluded-table}%
\end{table}%

\para{Question \ref{unremovable}:}
To test whether users were distracted by undesirable sequences that cannot be removed we first located undesirable sequences by how many users who had the ability (i.e. access to a $\exclude$ predicate) removed them, then counted their appearance in the sessions of users from the PBE group.
\Cref{excluded-table} shows the results. It is important to note that not all sequences that were commonly removed appeared in the PBE group at all, itself an indicator of how syntax operations vastly change the traversal of the search space.

These undesirable sequences appeared in user sessions up to $7$ times in a single session. Some of these sequences distracted (i.e. kept reappearing) up to $\nicefrac{2}{3}$ of the users performing a task, and on average $22.2$\% of the users.
Furthermore, a distracting sequence appeared, on average, around $3$ times in each session. This shows that the inability to remove a letter or sequence discussed in \cref{impossibility-claim} is not only a theoretical problem, nor is it only a problem at the end of the process as seen in \cref{all-sumup}, but a real distraction from the ability to synthesize over an expressive vocabulary.

%\TODO{XXX discuss histogram maybe}
When examining the presence of distracting elements in the final program accepted by participants, we can see in \cref{Fi:correctness} that in two of the tasks (histogram and nonemptylines) most or all PBE users ended up accepting a program with superfluous elements.
For example, many histogram sessions accepted a program with a call of \scode{toMap} on a map, and many nonemptylines sessions accepted a program that called \scode{filterNot(c => c == '\textbackslash r' || c == '\textbackslash n')} on a list of strings. Both of these are nullipotent elements: \scode{toMap} creates a map from a map, and \scode{filterNot(c => c == '\textbackslash r' || c == '\textbackslash n')} compares strings to characters and so always filters nothing.

%Users, in general, tried to eliminate these superfluous program elements. While some syntax-only and GIM users also failed to get rid of them, they were \emph{able to} -- in PBE the only way users could get to a program that is not equivalent is to select a program that handles fewer end-cases but we have decided to mark as correct.

In addition, in these cases where PBE users stopped at an equivalent program rather than the target program, we tested the number of iterations spent in the same equivalence class (i.e. presented with the same candidate program) before accepting the program.
While most users accepted an equivalent program immediately, one user performing the histogram task tried an additional iteration and one user tried two additional iterations. For nonemptylines, two users tried an additional iteration and one user tried two additional iteration. Altogether, these are $22\%$ of the sessions where
users tried unsuccessfully to improve upon the program they already had, either trying to get rid of a nullipotent element or not realizing it has no influence, before finally accepting it.

We chose not to tackle the questions of user preference and measures of distraction with a questionnaire asking the users to approximate their preference, sticking only to empirical results. Despite that, we wish to bring several anecdotes from the course of the experiment that may help shed light on the behavior observed. Users in the PBE test group expressed very specific frustration on several occasions such as \emph{``it insists on using \scode{take(5)} no matter what I do''} while solving the most frequent word task, or \emph{``I couldn't get rid of these nonsense functions, I just wanted to shake it''} after solving the non-empty lines task.

\begin{figure*}
%\begin{tabular}{\textwidth}{ccc}
\centering
\begin{minipage}[t]{.32\textwidth}
\includegraphics[width=\textwidth]{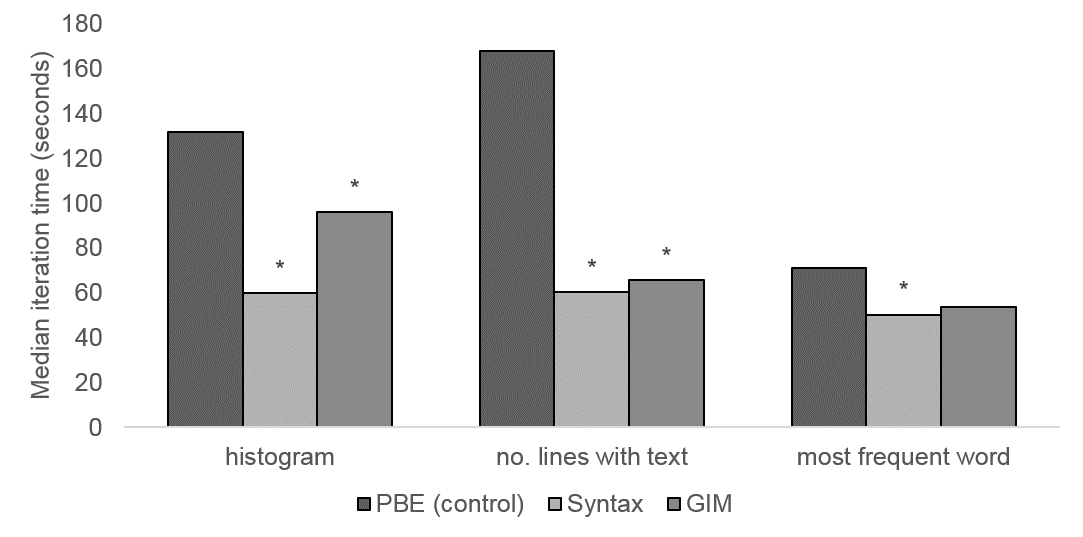}
\captionof{figure}{Median iteration time per task in each  test group.
Significant change from PBE indicated by *.}\figlabel{meditertime}
\end{minipage} \hfill%
\begin{minipage}[t]{.32\textwidth}
%  \centering
\includegraphics[width=\textwidth]{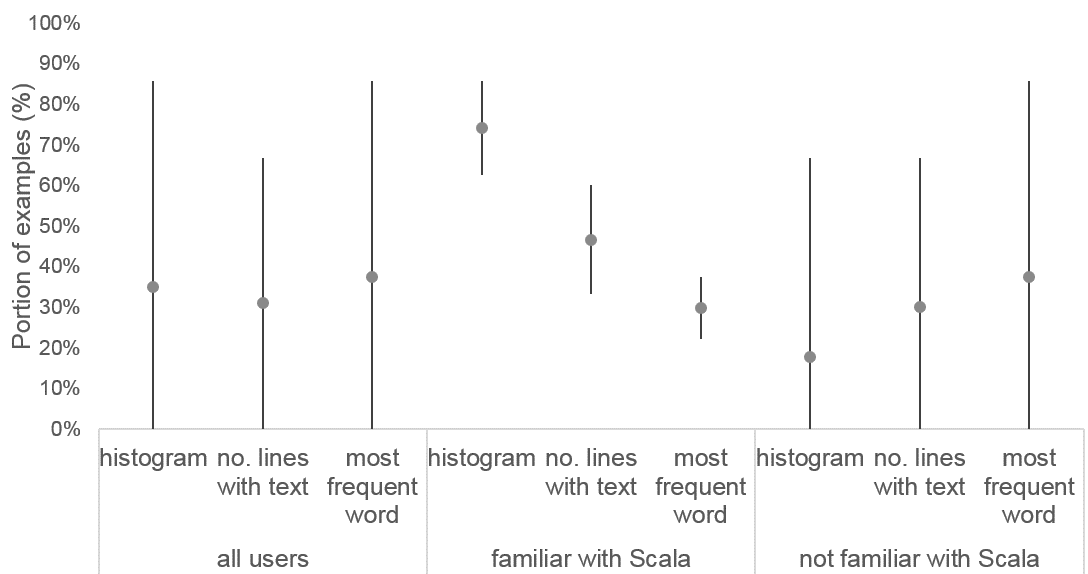}
\caption{Examples used (med, min, max) by GIM users (all operations). None used $100\%$ examples.}\figlabel{examples-minmaxmed}
\end{minipage} \hfill%
\begin{minipage}[t]{.323\textwidth}
\centering
\includegraphics[width=\textwidth]{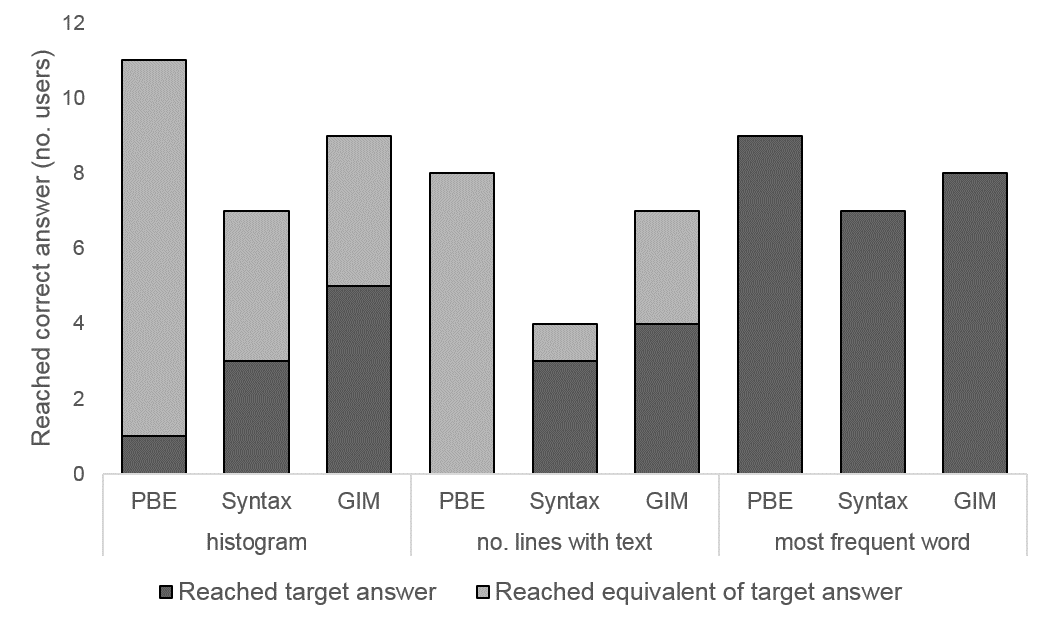}
%\captionof{figure}{Number of users that reached the correct result in each test, as well as those that reached an equivalent program with non-functional differences, such as a redundant functions constituting a null statement being part of the accepted solution.}\figlabel{correctness}
\captionof{figure}{No. users that reached the target program or an equivalent (non-functional differences) result.}\figlabel{correctness}
%\end{minipage}\hfill%
%\begin{minipage}[t]{.48\textwidth}
%\centering
%\includegraphics[width=0.95\textwidth]{stuck}
%\captionof{figure}{Number of iterations spent in PBE sessions without reducing the equivalence class before accepting the candidate program. This is indicative of trying to rule out a program equivalent to the target program but including superfluous operations.}\figlabel{stuck}
%\end{minipage}
\end{minipage}
%\end{tabular}
\end{figure*}

\subsection{Discussion and conclusions}\label{study-conclusions}

In this section we discuss the results of the study.

\para{Speed and ease of use}
We see a speedup of iteration time between examples and other predicates. The change is largest between the PBE and Syntax groups, and a smaller speedup when examining the GIM group against itself. We may attribute this difference between the two tests to the fact that the users in the GIM group resort to examples only when they are convenient or readily apparent and therefore take a shorter amount of time to create.

When combined with a low preference for examples, we conclude that syntax predicates are \emph{easier} for the user to use in general.

In addition, when we combine the improvement in iteration time, the change in number of iterations (itself statistically significant) and the lack of significance in the change in total time, we conclude that changing the type of predicates simply leads to the same time spent on the synthesis task \emph{using more, but shorter and easier, iterations}.

\para{Distracting elements and user frustration}
Much of the frustration users in the PBE group expressed had to do with recurring program elements they thought were useless. Recurring undesirable sequences showed up in up to \nicefrac{2}{3} of the users and recurred on average $3$ times during the session, which is definitely a reason for frustration.
In addition, some PBE users wasted time and effort trying to remove elements that cannot be removed.
We therefore conclude that avoiding this distraction by giving users more tools would, at the very least, make for more content users.

\para{Helpfulness of debug information}
We attribute the success rate and relatively short use times of a set of developers who have never before seen Scala to the guidance offered by debug information. We did not target this specifically in the experiment, but $9$ separate users volunteered to us after the experiment that it was anywhere from ``helpful'' to ``lifesaving'' in understanding unfamiliar methods and keeping track of examples. This approval included people who were familiar with Scala and developers who develop in Scala.

\para{Correctness with syntax operations}
There is a dip in correctness (a functionally correct program was not reached) from the PBE group to the Syntax group, but an improvement in GIM. We attribute this to the helpfulness of debug information: it seems to be easier to make a correct decision about a program when presented with its breakdown over several examples, rather than just the single initial example available to the Syntax group.

\para{Preferred operations}
There is a very strong preference of all users for syntactic predicates over examples. However, there may be a subordinate trend within this preference:
separating the users into those familiar with Scala and those who are not, the preference reverses. Users familiar with Scala preferred more examples than the rest, and preferred examples over other predicates in the harder task, histogram, and predicates over examples in the easier task, frequword. This may have to do with their ability to better understand a candidate program: more savvy programmers could more easily read the programs and so preferred to break the observed behavior with examples, whereas programmers that have a harder time reading the code focus on individual program elements. This remains a conjecture as the set of users familiar with Scala within the GIM group is only $2$.

\section{Threats to Validity}

%Internal
\para{Cross validation} The study was not cross-validated (i.e. each user performing tasks in several groups). Because the predicate families include each other, we felt it would create a bias toward some operations based on order. As cross validation should not be used when it creates bias, we decided against it.
We tried to negate some of the differences between individual programmers tested
%This leaves differences between individual programmers tested, which we tried to negate
by drawing participants from similar backgrounds -- same year in university, developers in the same department -- and then dividing them evenly.

%External
\ignore{
\para{Drop-outs}
Two of the $32$ participants did not complete all tasks, both from the Syntax group. This introduces two problems: first, since the users in question only performed one or two sessions, these are sessions where they are still less adept. Second, they create an imbalance in the test groups when comparing the exercises after they dropped out. However, since the impacted group was the Syntax group, most conclusions about the GIM test group are unaffected.
}

\para{Sampling of population}
%(a) low percentage of scala programmers,
An external validity issue mentioned in \cref{study-conclusions} is the relatively small percentage of participants familiar with Scala -- only $25\%$ of the participants of the study, and, due to random assignment to groups, in some of the groups as few as $20\%$. As we have already pointed out, this does not allow us to make general claims about differences between Scala-savvy programmers and those who are new to the language. However, we can still generalize our claims with regards to programmers working in a language they have not encountered before -- the majority of the participants.
%(b) generally not a ``sample''
Additionally, our sample of undergraduates is not random, but rather of students who felt familiar enough with functional programming to agree to participate. This may skew the ability to generalize. This is hopefully not significant in the compiled results as undergraduates are not a very large percentage of the participants -- less than $22\%$. 
\section{Related Work}

\para{Syntax-based synthesis} \cite{alur2015syntax} is the domain of program synthesis where the target program is derived from a target programming language according to the syntax rules. \cite{itzhaky2016deriving,lau2001learning,udupa2013transit,raychev2014code} all fall within this scope. The implementation of GIM presented in this paper is syntax-based, where the target language is a functional subset of Scala as specified by $\vocab$. Syntax-based synthesis algorithms often use a \emph{user-driven interaction model}~\cite{gulwani2012synthesis} which GIM extends.

%PBE-Userdriven
\para{Programming by Example} In PBE the interaction between user and synthesizer for demonstrating the desired behavior is restricted to examples, both in initial specifications and any refinement.
%FlashFill
FlashFill~\cite{Gulwani:2011:ASP:1926385.1926423,polozov2015flashmeta} is a PBE tool for automating transformations on an Excel data set, and is included in Microsoft Excel. It does not show its users the program, only its application on the data set. %This means that the result of the synthesis is not reusable to any other data set since it might hide unintended behavior that happens to work on the current data set.
Because the resulting program is never inspected, it might still suffer from overfitting to the examples and is not reusable.
%PBE-Userdriven+shows code????
Escher~\cite{albarghouthi2013recursive} is a PBE tool for synthesizing recursive functions. Like FlashFill, Escher decomposes the synthesis task based on the examples, searching for programs that could be used as sub-programs in condition blocks.
Escher is parameterized by the set of operations used in synthesis, and
%does not allow the user to contribute feedback as the algorithm progresses.
like FlashFill, allows refinement only by re-running the process.

%TypeDriven (User-driven)
\para{Type-Directed Synthesis} is a category of synthesis algorithms that perform syntax-based synthesis mainly driven by the types of variables and methods, and the construction of the program is performed through type-derivation rules. While type-directed methods tend to be user-driven, many of them~\cite{gvero2013complete,galenson2014codehint,perelman2012type} require only initial specifications and the user manually chooses from multiple candidate programs that match the specification. The philosophy behind GIM is that a user shouldn't consider many programs (could be dozens or more) at a time with no additional data. Rather, programs should be considered one at a time, with additional information that can help the user consider the program in depth and direct the search.

%Mixing PBE with TypeDriven
\para{Adding examples to Type-directed synthesis} Recent work connects PBE with type-driven synthesis~\cite{osera2015type,feser2015synthesizing}. These tools accept their initial specifications as examples (and their inherent type information), use type derivations to produce candidate programs, and verify them with the examples.
%M/R
{\small \textsc{Big$\lambda$}}~\cite{smith2016mapreduce} synthesizes MapReduce processes via sketching and type derivations over lambda calculus and a vocabulary. Examples are also used to verify determinism. %Examples are used for both type information and for verification, and refinement is performed by adding more examples. %If the program that was returned by {\small \textsc{Big$\lambda$}} is rejected by the user, they will add more examples (leading, again, to the problem of communicating intent through examples) and re-run the tool.
%Petrinet
{\small \textsc{SyPet}}~\cite{petrinetsynth17} is a type-directed, component-based synthesis algorithm that uses  Petri-nets to represent type relationships, 
and finds possible programs by reachability. Candidates are tested using tests provided by the user. %In addition to the fact that {\small \textsc{SyPet}} accepts no user feedback, it also
{\small \textsc{SyPet}} requires full test cases rather than examples, which, while more descriptive, still require the user to learn a lot about the library in order to program the test case, effort that may be equal to learning about the methods required to solve the programming task at hand.

\para{Sketching}
The user can restrict the search space via sketches~\cite{solar2008program,solar2006combinatorial,solar2008sketching}, structural elements (e.g. conditions or loops) which includes holes to be synthesized. Sketching is a way to leverage a programmer's knowledge of expected syntactic elements, and when used in conjunction with restrictions on the syntax~\cite{alur2015syntax} can allow very intricate synthesis. However, since the most general sketch, a program with only a single hole, is usually too unconstrained for the synthesizer, the user must come armed with at least some knowledge of the expected structure rather than iteratively build it as in GIM.

\para{Enriching user input} Several existing works have enriched the specification language, or the interface for specifying program behavior. Adding examples to type-directed synthesis is an example of such enrichment. Another approach by Polikarpova et al.~\cite{polikarpova2016program} with {\small \textsc{Synquid}} is to use refinement types instead of types, which encode constraints on the solution program, which can be imposed on the candidate space. While these constraints are mainly semantic, unlike GIM's syntactic predicates, this embodies the same ideal of passing off some responsibility to a user that can understand code, or in this case, write code.
Likewise, Barman et al.~\cite{barman2015toward} suggest an interactive, user-dependent extension of sketching intended to synthesize \emph{the sketch} itself by leveraging the user to decompose the specifications and examine the results. Angelic programming~\cite{bodik2010programming} leverages programmer knowledge by an expanded interface from synthesizer to user: the user is shown a synthesized program with a nondeterministic ``angelic operation'' and execution traces for that operation to make the program correct, and the it is their responsibility to identify the needed operation to replace the angelic operator.

\section{Conclusion}

We presented a novel granular interaction model (GIM) for interacting with a synthesizer. This interaction model extends common PBE approaches and enables a programmer to communicate more effectively with the synthesizer.

First, we prove that using only examples is insufficient for eliminating certain undesired operations in a program, where these undesired operations are easy to eliminate when using syntactic operations made available by GIM.

Second, we show the effectiveness of GIM by a controlled user study that compares GIM to standard PBE. Our study shows that participants have \emph{strong preference} (66\% of the time) to using granular feedback instead of examples, and are able to provide granular feedback up to 3 times faster (and 2.14 times faster on average).

\section{Acknowledgements}
The research leading to these results has received funding from the European Union's - Seventh Framework Programme (FP7) under grant agreement no. 615688 - ERC- COG-PRIME.

The authors thank Yifat Chen Solomon for her indispensable assistance in getting the user study off the ground, and Hadas E. Sloin and Yoav Goldberg for their help in analyzing the data.

\begingroup
\let\clearpage\relax

\bibliographystyle{ACM-Reference-Format}
\bibliography{bib}
\endgroup

\end{document}